\documentclass[useAMS,usenatbib,usegraphicx]{mn2e}
\usepackage{latexsym}
\usepackage{amsmath}
\usepackage{amssymb}
\def\ltsima{$\; \buildrel < \over \sim \;$}
\def\lsim{\lower.5ex\hbox{\ltsima}}
\def\Msunh{\mbox{$h^{-1}$M$_\odot$}}
\def\mpch{\mbox{$h^{-1}$Mpc}}

\def\deg{\ifmmode{^\circ}\else{$^\circ$}\fi}
\def\hGpc{\ifmmode{h^{-1}{\rm Gpc}}\else{$h^{-1}{\rm Gpc}$}\fi}
\def\hkpc{\ifmmode{h^{-1}{\rm kpc}}\else{$h^{-1}{\rm kpc}$}\fi}
\def\hMpc{\ifmmode{h^{-1}{\rm Mpc}}\else{$h^{-1}{\rm Mpc}$}\fi}
\def\hMsun{\ifmmode{h^{-1}M_\odot}\else{$h^{-1}M_\odot$}\fi}

\def\muK{\ifmmode{\mu{\rm{K}}}\else{$\mu$K}\fi}
\def\mum{\ifmmode{\mu{\rm{m}}}\else{$\mu$m}\fi}

\def\sige{\mbox{$\sigma_8$}}

\newcommand{\LCDM}{$\Lambda$CDM}
\newcommand{\aj}{{AJ}}
\newcommand{\apj}{{ApJ}}
\newcommand{\apjl}{{ApJ}}

\newcommand{\mnras}{{MNRAS}}

\title{The statistics of voids as a tool to constrain cosmological parameters: $\sigma_{8}$ and $\Omega_{m}h$}

\author[Betancort-Rijo et al.]{
\parbox[t]{\textwidth}{Juan Betancort-Rijo$^{1,2}$\thanks{E-mail:jbetanco@iac.es},
Santiago G. Patiri$^{3}$\thanks{E-mail:
spatiri@case.edu},
Francisco Prada$^{4}$,\\  
Antonio Enea Romano$^{5}$}
\\
\\
$^1$
Instituto de Astrofisica de Canarias,
C/ Via Lactea s/n, Tenerife, E38200, Spain
\\
$^2$
Facultad de Fisica, Universidad de La Laguna, Spain
\\
$^3$
Department of Astronomy,
Case Western Reserve University,
Cleveland, OH 44106, USA
\\
$^4$
Instituto de Astrofisica de Andalucia (CSIC), E-18008, Granada, Spain
\\
$^5$
Yukawa Institute for Theoretical Physics, Kyoto University,
Kyoto 606-8502, Japan
\\
}

\begin{document}

\maketitle

\begin{abstract}

We present a general analytical formalism to calculate accurately several statistics 
related to underdense regions in the Universe. The statistics are computed for dark matter halo and galaxy distributions 
both in real space and redshift space at any redshift. 
Using this formalism, we found that void statistics for galaxy distributions can be obtained, 
to a very good approximation, assuming galaxies to have the same clustering properties as halos above a certain mass. 
We deducted a relationship between this mass and that of halos with the same accumulated number density as the galaxies. 

We also found that the dependence of void statistics on redshift is small. For instance, the number of voids larger 
than $13 \mpch$ (defined to not contain galaxies brighter than $M_{r}=-20.4 +5log h$) change less than 20\% between $z=1$ and $z=0$. 
However, the dependence of void statistics on $\sigma_{8}$ and $\Omega_{m}h$ is considerably larger, making them appropriate to develop 
tests to measure these parameters. We have shown how to efficiently construct several of these tests and discussed in detail the 
treatment of several observational effects. The formalism presented here along with the observed statistics extracted 
from current and future large galaxy redshift surveys will provide an independent measurement of the relevant cosmological parameters. 
Combining these measurements with those found using other methods will contribute to reduce their uncertainties.

\end{abstract}


\section{Introduction}

Large underdense regions in the Universe, commonly known as voids, are a very relevant signature of the large-scale structure of the Universe (LSS).
Not until recently have they drawn too much attention from both observational and theoretical sides. However, it is becoming more evident that voids
play an important role as cosmological probes and in understanding the processes involved in galaxy formation 
(Peebles 2001; Plionis \& Basilakos 2002; Hoyle et al. 2005; Conroy et al. 2005; Patiri et al. 2006a,b; Tinker, Weinberg \& Warren 2006; Hoeft et al. 2006; Park \& Lee 2007; Croton \& Farrar 2008; Tinker et al. 2008). 

Voids occupy large areas projected into the sky, and only recently galaxy redshift surveys have become large enough
to allow systematic and robust studies of void statistics (see e.g. Croton et al. 2004; Hoyle \& Vogeley 2004; Patiri et al. 2006; Ceccarelli et al. 2006 for analysis
of voids in the 2dF and Tinker et al. 2008 in the SDSS).
On the theoretical side, important progress has been made recently by improving analytical and numerical simulation modelling of voids (Mathis \& White 2002;
Gottl\"ober et al. 2003; Colberg et al. 2005 for numerical simulations,  Sheth \& Van de Weygaert 2004; Patiri, Betancort-Rijo \& Prada 2006,
hereafter PBP06, for analytical approaches).

Most of the recent efforts concerning the statistics of voids have been focused on constraining different aspects of the formation and evolution of structure in the Universe rather than on assessing their 
ability to restrict cosmological parameters. The statistic of voids can, in principle, be used to put constraints to any cosmological parameters. However, these statistics may be  
especially sensitive to some of them.
For instance, the abundance of voids larger than a given radius depends on the normalization of the linear spectrum of mass fluctuations (noted as $\sige$).
Furthermore, the abundance of voids as a function of their radius depends on the shape of the spectrum (usually denoted as $\Gamma \equiv \Omega_{m}h $). Although these dependencies arise naturally from the current picture of structure formation, 
we must note that it has been difficult to prove them (see Little \& Weinberg 1994; Tinker, Weinberg \& Warren 2006). 
We argue that this might be due to the specific way how the galaxies are distributed within dark matter halos. In previous works, Halo Occupation Distribution (HOD) and its variants have been used to model de distribution of galaxies within halos. In these methods, the models are fitted (for a given $\sigma_{8}$, for instance) to reproduce the projected two-point correlation function, which renders very similar void statistics, resulting in an insensitivity of the void statistics with $\sigma_{8}$ (e.g. Tinker, Weinberg \& Warren 2006). Also, the 
limited size of simulation boxes used in previous works could be an issue. Note that the most widely used void statistics to test these assumptions has been the so-called Void Probability Function (VPF). However, it has large sampling errors, and may be not the most efficient statistic, being other statistics like the number of voids 
{\it larger} than a given radius a more appropriate approach (Plionis \& Basilakos 2002; PBP06). 

Computing accurate predictions of void statistics for different cosmological parameters is not a simple task. The options are running a large suite of numerical simulations, for different sets of cosmological parameters, with enough volume to obtain a reliable void statistics or develop an analytical framework to predict the dependence of 
statistics of voids with cosmological parameters. As the former involves 
a major consumption of computing power in order to achieve the needed accuracy, in this paper we focus our effort on the latter approach. In PBP06 we already developed an analytical framework to obtain the number densities of voids larger than a given radius defined by dark matter halos in a \LCDM~ cosmology.
Based on our previous work, we present here an extension to that formalism in order to compare model predictions with observational void statistics in a homogeneous way, and consequently, be able to extract cosmological information.
The extension we refer to is done in two aspects. First of all, the relationship between the VPF and the number density of voids, which was determined in PBP06 only for the largest void limit (for a given number density of dark matter halos), is extended here to deal also with smaller voids. In addition, the analytical expression to compute the VPF of dark matter halos itself was improved. Secondly, we modified our formalism so that we can now compute directly the predictions for the number of voids larger than a given radius {\it observed} within a given volume (i.e. taking into account the redshift distortions and the selection effects of a given survey).

Currently, there are several methods and experiments devoted to constrain cosmological parameters. CMB experiments such as WMAP (Hinshaw et al. 2008; Komatsu et al. 2008) 
in combination with galaxy clustering information (e.g. Sanchez et al. 2006; Seljak et al. 2006) have provided the best constrains to date. Methods based on clusters of galaxies and 
lensing are also promising (e.g. Bahcall et al. 2003; Wang et al. 2003; Yoo et al. 2006; Vikhlinin et al. 2008). However, as all methods have limitations and even degeneracies 
in some parameters, it is essential to provide independent parameter estimations in order to improve the accuracy of the combined measurements. As mentioned above, voids might be specially interesting to measure some cosmological parameters, such as the normalization of the amplitude of density fluctuations since they are tracing comparable scales.  

The necessary steps to achieve the goal of using the statistics of voids to constrain cosmological parameters are presented as follows: in section 2 we present the new relationship between the number density of voids [$\bar{n}(r)$] and the VPF [denoted as $P_0(r)$]. In section 3 we show how to obtain the VPF and other statistics for galaxy distributions, discussing the relationship between dark matter halos and galaxies. In section 4 we extend our formalism in order to compute voids statistics in redshift space. In section 5 we discuss the dependence of voids statistics on redshift and cosmological parameters ($\sigma_8$ and $\Gamma$ in particular). In section 6 we show how to handle in the theoretical calculations several observational effects that perturb the statistics of voids, namely the effect of the finite sample, the spectroscopic completeness and the variations of the void statistics with redshift (i.e. the {\it snapshot} effect).  In section 7 we propose different tests 
for measuring cosmological parameters using void statistics. In section 8 we present a comparison of results obtained using our formalism with those found in numerical simulations (Millennium Run). Finally, in section 9 we present the discussion and conclusions.

\section{Relationship between the VPF and the number density of voids}

The most common, widely used void statistic is the so-called Void Probability Function (VPF, White 1979), which is the probability 
that a randomly placed sphere with radius $r$ is empty of objects (galaxies or dark matter halos). Another important statistic is the 
number density of voids (defined as maximal non-overlapping spheres) larger than a given radius $r$. Several works have established the relationship 
between these two statistics (see e.g. Otto et al. 1986; Betancort-Rijo 1990; PBP06). However, they are only valid for rare voids, i.e. the largest voids in a given sample.

In PBP06 we argued that most of the information carried by void statistics, concerning cosmological parameters, comes from rare voids.
However, as we will see below, more common voids are still relevant to increase the statistics, which is fundamental to constrain cosmological parameters reliably. 
In order to take into account more common voids, the analytical relationships mentioned above have to be improved.

In PBP06 we showed that the VPF [that we denote here as $P_{0}(r)$] is related to the number density of voids larger 
than a radius $r$ [$\bar{n}_{v}(r)$] by the following expression (for the rare voids limit):

\begin{equation}
\bar{n}_{v}(r) \simeq \frac{3\pi^2}{32}\frac{(\bar{n}'V)^3}{V}P_{0}(r) \label{eq:eq1}
\end{equation}
where
\begin{equation}
\bar{n}'V=-\frac{1}{3}\frac{d\ln P_{0}(r)}{d\ln r} ;\qquad V=\frac{4}{3}\pi r^3   \label{eq:eq2} 
\end{equation}
and $\bar{n}'V$ is the mean density of points in the surface of a randomly chosen empty sphere with radius $r$, that we denote in term of 
the derivative of $P_{0}(r)$ with respect to $r$.

Equation (\ref{eq:eq1}) gives an unique functional relationship between $P_{0}(r)$ and $\bar{n}_{v}(r)$. As mentioned above, this works well for 
rare voids, but for more common voids the existence of a unique functional relationship has to be studied. Note that the VPF is determined by all the 
hierarchy of correlations functions (White 1979), but the VPF itself do not determine uniquely all these correlations. For instance, two samples could have 
the same VPF and still differ in some aspect of the clustering, which will render different number densities of voids larger than a given radius.
To address this issue we carried out a detailed analysis using the Millennium Run numerical simulation (Springel et al. 2005; see section \ref{res}). We found evidence for an unique 
functional form of the VPF for the distributions relevant to this work. We write this expression as an extension of the relationship given in equation (\ref{eq:eq1}), i.e.

\begin{equation}
\bar{n}_{v}(r) \simeq \frac{0.68K(r)}{V} e^{-3.5K(r)[1-2.18K(r)]} \label{eq:eq3}
\end{equation}
and
\begin{equation}
K(r)=\bigg(-\frac{1}{3}\frac{d\ln P_{0}(r)}{d\ln r}\bigg)^3 P_{0}(r) \quad;\quad V=\frac{4}{3}\pi r^3 \label{eq:eq4}
\end{equation}
These equations are valid for $K(r) \leq 0.46$, while for $K(r)> 0.46$ $\bar{n}_{v}(r)=0.313/V$. The quantity $K(r)$ measures the rareness of the voids. 
In the rare void limit $K(r)$ goes to zero, so the exponential factor is close to one, recovering the original equation (\ref{eq:eq1}). Note also that 
the coefficient $3\pi^2/32$ shown in equation (\ref{eq:eq1}) is replaced by 0.68 in equation (\ref{eq:eq3}). The former coefficient was originally introduced 
by Preskill \& Politzer (1986), but we found the latter to be more accurate (Betancort-Rijo, in preparation). 

In the limit in which $K(r)$ goes to zero, all voids larger than $r$ are only slightly larger than $r$, so that their mean volume, $\bar{V}(r)$, is only 
slightly larger than $V(r)$. Thus, using equation (\ref{eq:eq4}) we have for the fraction of volume occupied by voids, $F(r)$:
 
\begin{equation}
F(r)=\bar{n}_{v}(r)\bar{V}(r) \gtrsim \bar{n}_{v}(r)V(r) \simeq 0.68K(r) \label{eq:eq5}
\end{equation}
\\
For $K(r)$ less than 0.1, $F(r)$ is smaller than 0.07 and so the voids can be considered rare. For these voids  exponential factor in equation (\ref{eq:eq3}) differs 
less than a 10\% from 1 so that the asymptotic expression [eq. (\ref{eq:eq1}) with the adequate coefficients] is a good approximation. As $K(r)$ increases, 
the exponential factor decreases, reaching a minimum at $K(r) \simeq 0.23$. Here, the exponential part takes a value close to 2/3. For $K(r) > 0.23$, the exponential factor increases again, 
recovering the value of 1 for $K(r) \simeq 0.46$. Note that this corresponds to rather common voids occupying more than one third  of the volume of the sample. Further, for very 
commom voids $\bar{n}_{v}(r)$ is practically independent of $K(r)$. In this limit the problem degenerates into a problem of random packing of spheres of unequal 
radius (see e.g. Shi \& Zhang 2008), which is far beyond the scope of this work (it is worth to mention, however, that equation (\ref{eq:eq3}) provides a reasonably good approximation even in this limit, although it have carefully been checked only for $K \leq 0.42$.) 

To compute $P_{0}(r)$ we start from a more general statistic that is $P_{n}(r)$. This denotes the probability that a sphere of radius $r$, placed at random within the distribution, contains $n$ objects (Layzer 1954). Note that for small values of $n$, $P_{n}(r)$ is still characterizing underdense regions. As we will see, this statistic is very important for our purposes. For point distributions conforming to a random, non-uniform Poissonian process (Peebles 1980) we have that 
 
\begin{equation}
P_{n}(r)=\int_{0}^{\infty}P(u) \frac{u^n}{n!}e^{(-u)} du   \label{eq:eq6}
\end{equation}
where $P(u)$ is the probability distribution for the integral of the probability density, $u$, within a randomly placed sphere. In PBP06 we showed that for dark matter halos, $u$ can be written as:

\begin{equation} 
u=[\bar{n} V (1+\delta)][1+\delta_{ns}]  \label{eq:eq7}
\end{equation}
where $\bar{n}$ denotes the mean number density of those halos in the sample (usually halos larger than some given mass). $V$ is the 
volume of the sphere. $\delta$ is the actual enclosed density contrast within the sphere. The first term in the right hand side of the equation is the 
integral of the probability density within the sphere for halos tracing the mass (i.e. no bias, which is true in the very low mass limit). In general, 
halos are biased tracers of the underlying mass distribution, due to the initial clustering of the proto-halos before they move along with mass (i.e. the statistical clustering). 
The second term of the equation accounts for this biasing. In PBP06 we obtained an approximation for this bias as a function of the linear enclosed density contrast 
within the sphere ($\delta_{l}$) written as:
 
\begin{equation} 
1+\delta_{ns}(\delta_{l})=A(m)e^{-b(m)\delta_{l}^2} \quad \forall \quad \delta_{l} \leq -1 \label{eq:eq8}
\end{equation}
where $A(m)$, $b(m)$ are coefficients mainly depending on the halo mass, and to a lesser extent, on the size of the sphere (see Rubi\~no-Martin et al. 2008). 
We use the complete Zeldovich approximation (CZA, Betancort-Rijo \& L\'opez-Corredoira 2001) to obtain the actual enclosed density contrast ($\delta$), as a function 
of the eigenvalues of the linear deformation tensor, $\lambda$, and since $\delta_{l}=\sum \lambda_i$, we can write $u$ as a function of the $\lambda$ values. 
Equation (\ref{eq:eq6}) can be then written as:
  
\begin{eqnarray}
P_{n}(r)=\int\int\int P(\lambda_1,\lambda_2,\lambda_3,r) \frac{[u(\lambda_i)]^n}{n!} \times \nonumber\\
\times e^{[-u(\lambda_i)]} d\lambda_1d\lambda_2d\lambda_3  \label{eq:eq9}
\end{eqnarray} 
where $P(\lambda_{i},r)$ is the probability distribution for the $\lambda_{i}$ within a sphere of radius $r$ chosen at random in Eulerian space (more precisely, the $\lambda$ values on the Lagrangian patches that transform into that sphere). Betancort-Rijo \& Lopez-Corredoira (2002) showed that $P(\lambda_i,r)$ can be derived 
from the probability distribution for the $\lambda$ within a sphere of constant Lagrangian radius $Q$ chosen at random in Lagrangian space (Doroshkevich 1970). 
Note that in this last equation a triple integral is involved. However, in the cases we are interested in [i.e. where $n$ is rather smaller than the mean $(\bar{n}V)$], 
this equation can be approximated without loss of accuracy by an equation involving only one integral:
  
\begin{equation}
P_{n}(r)=\int_{-\infty}^{1.6} P(\delta_l,r) \frac{[u(\delta_l)]^n}{n!}e^{[-u(\delta_l)]} d\delta_l   \label{eq:eq10}
\end{equation}  
where $u(\delta_l)$ is now a function of $\delta_{l}$ (and implicitly of $r$) through the dependence of the actual density contrast, $\delta$, on its linear 
counterpart, $\delta_l$. We write this functional dependence as
  
\begin{equation} 
u(\delta_l)=(\bar{n}V[1+\delta(\delta_l,r)])[1+\delta_{ns}(\delta_l)].   \label{eq:eq11}
\end{equation}
$\delta(\delta_l,r)$ is basically the relationship between the actual and the linear density contrast within a sphere as given by the standard spherical collapse 
model, except for a small correcting term depending on $r$ (see eq. A3 in Appendix A). Note that in PBP06 we did not use this correcting term, so the values of $P(r)$ obtained 
there have a small but non-negligible error. $P(\delta_{l},r)$ is the probability distribution for the linear density contrast within a sphere chosen at random in 
Eulerian space. See Appendix A for details on how to efficiently evaluate equation (\ref{eq:eq10}).
  
In PBP06 we argued whether halo clustering conform exactly to a random Poissonian model. In this model, objects are placed in the distribution accordingly 
to an underlying probability density field, but independently of the actual position of the placed points. In fact, halos have an exclusion region around them that 
can not be accounted for within that model. However, we do not detect a deviation from the Poissonian model. This is probably due to the fact that at least in these underdense environments the mean distance between halos is much larger than the mentioned exclusion region (see also Conroy et al. 2005).

\section{Voids in galaxy distributions}

In the previous section we described how to compute predictions for the number density of voids defined by dark matter halos. However, we do observe galaxies, which are assumed to be embedded in those dark matter halos. Hence, we need to establish a relationship between them in order to obtain $P_{n}(r)$ for galaxy distributions and then be able to compare model predictions with observations. 

From equation (\ref{eq:eq10}) we see that the relationship between halos and galaxies enters only through $\delta_{ns}$, which quantifies the biasing of the galaxies with respect to the underlying matter distribution. We denote the function carrying the biasing for galaxies as $\delta_{Ls}$. This function is determined by the $\delta_{ns}$ for halos through the relationship of galaxies with halos. The general equation to compute $P_{n}(r)$ for a galaxy distribution can be written in a similar way than equation (\ref{eq:eq10}), but replacing $\delta_{ns}$ by $\delta_{Ls}$. In PBP06 we found (see also Yang et al. 2003) that 

\begin{equation}
1+\delta_{Ls}(\delta_{l},L)=\frac{\int_{0}^{\infty}\Phi(>L|m,\delta_{l})n_{c}(m,\delta_{l})~dm}{\Phi_{u}(>L)}. \label{eq:eq12}
\end{equation}
Here, $n_{c}(m,\delta_{l})$ is the conditional mass function within a region with linear density fluctuation $\delta_{l}$. $\Phi(>L|m,\delta_{l})$ is the cumulative  conditional luminosity function and $\Phi_{u}(>L)$ is the unconditional one. It is worth to mention that in this equation we allow for a dependence of the conditional  luminosity on the environmental density through $\delta_{l}$. However, there is an increasing evidence favoring a conditional luminosity not depending on environment for a large range of luminosities (see Tinker et al. 2008; Tinker \& Conroy 2008).

Given the conditional luminosity function, $\Phi(>L|m,\delta_{l})$, equation (\ref{eq:eq12}) can be used to obtain $\delta_{Ls}$ as a function of $\delta_{l}$. However, this is neither simple nor straightforward. Here, we prefer to determine a functional form from general considerations and calibrate it with numerical simulations. We know that for halos above any given mass the dependence of $1 + \delta_{ns}$ on $\delta_{l}$ is accurately fitted by a Gaussian (equation [\ref{eq:eq8}]). This implies that $1+\delta_{ns}$ have to be well approximated by a Gaussian for halos with mass within any mass interval. Now, in equation (\ref{eq:eq12}) we see that $1+\delta_{Ls}$ is the average of the value of $1+\delta_{ns}$ over all halo masses (weighted by the probability distribution for the mass of a halo containing a galaxy with luminosity larger than $L$). Therefore, $1+\delta_{Ls}$ must be approximated by a Gaussian with an accuracy comparable to the accuracy of the approximation of $1+\delta_{Ls}$ by Gaussians within the relevant range of mass values. Then we can write:

\begin{eqnarray}
1 + \delta_{Ls} =& A(L)~e^{-b(L) \delta_{l}^{2}} ; \label{eq:eq13}\\ 
b(L) \equiv& b(m_{g}) \\
A(L) \equiv& A(m_{g})
\end{eqnarray}
where $b(m_{g})$, $A(m_{g})$ are the same functions $b(m), A(m)$ defined in equation (\ref{eq:eq8}) evaluated at mass $m_{g}$. Equation (\ref{eq:eq13}) means that the biasing of galaxies with luminosity larger than $L$ with respect to the underlying mass distribution is equal to the biasing 
of the halos with mass larger than $m_{g}$. Let $m$ be the mass such that the number density of all halos with mass larger than $m$ is equal to the number density of the galaxies under consideration, $\bar{n}$, i.e. $\bar{n}(>m)=\bar{n}_{g}$ [where $n(>m)$ is the cumulative cosmic mass function]. The relationship between $m_{g}$ and $m$ is given by:

\begin{equation}
m_{g} = m (1.396)^{\sigma(m)}  \label{eq:eq14}
\end{equation}
where $\sigma(m)$ is the {\it rms} linear density fluctuation on scale $m$. For very large masses $\sigma(m)< 1$ and $m_{g}$ is very close to $m$, but for masses with $\sigma(m)> 1$, $m_{g}$ may be substantially larger than $m$. Note that $m_{g}$ is not the mass (lower limit) of the halos containing the galaxies under consideration, but  the mass such that the clustering properties of the galaxies is equal to that of all halos more massive than $m_{g}$.

The form of equation (\ref{eq:eq14}), where $m_{g}/m$ depends only on $\sigma(m)$ is what is expected if the clustering of the galaxies under consideration is determined only by the process of gravitational hierarchical collapse. The specifics of the galaxy formation processes are not relevant in as much as equation (\ref{eq:eq13}) is a good approximation. The number density of galaxies of a given type above a given luminosity does strongly depends on those specifics, but the relationship between the number density (which gives m) and the clustering (given by $m_{g}$) does not.

Equation (\ref{eq:eq13}) is the result of fitting a functional form derived from general considerations to the results of the numerical simulations described in section \ref{res}. It must be noted, however, that the values of $\sigma$ used in fitting equation (\ref{eq:eq13}) only goes from 1.6 to 2.6, which is enough for the masses and redshifts relevant to our present problem. Further tests must be imposed on this equation before extrapolating it far beyond the well checked range of $\sigma$ values.

In Summary, to obtain the voids statistics for galaxies brighter than a given luminosity, we use equation (\ref{eq:eq10}) with $1+\delta_{ns}$ equal to that of halos 
larger than a certain mass $m_g$, which is determined by the number density of the galaxies.

\section{Voids in redshift space}

The equation given in the previous section for $P_{n}(r)$ corresponds to voids in real space. However, galaxy positions provided by redshift surveys are in 
redshift space. In Patiri et al. (2006) we introduced a method based on the standard spherical collapse model to transform to real space, one-by-one, the voids found in the 2dF Galaxy Redshift Survey. In the present work we will use a similar method. However, instead of transforming the observed statistic we will modify our equations in order to compute the model predictions directly in redshift space, which is more straightforward.

In the spherical expansion model the peculiar velocity of matter at distance $r$ from the center of a sphere with actual enclosed density contrast $\delta(r)$ is 
given by:

\begin{equation} 
V(r)=H~r~ {\rm VEL}[\delta(r)] \label{eq:eq15}
\end{equation}
where $H$ is Hubble constant and ${\rm VEL}[\delta(r)]$ is a unique function of the enclosed density contrast (for a specific cosmology, see Appendix A). 
The matter distribution within a maximal sphere (i.e., our definition of voids) is not exactly spherically symmetric, and outside that sphere the distribution 
of matter in scales much smaller and much larger than the radius is strongly non-spherical. However, the average velocity field around that sphere [over all 
voids with given $r$, $\delta(r)$] is described well by the standard spherical collapse model. This mean velocity field play a key role in 
transforming the statistics under consideration from real to redshift space. 
 
A sphere with radius $r$ in real space and with inner mean fractional density $\delta$ has a mean outflow in its surface given by equation (\ref{eq:eq15}). 
In redshift space, that sphere transforms into a prolate spheroid elongated along the line of sight. The semiaxes along that direction in redshift space, $r^{\star}_{\parallel}$, 
is given by: 

\begin{equation}
r^{\star}_{\parallel}=r(1+{\rm VEL}[\delta(r)]) \label{eq:eq16}
\end{equation} 

However, the transverse axes $r^{\star}_{\perp}$ remain equal to $r$. This result is exact for any value of ${\rm VEL}[\delta(r)]$. Conversely, for small values 
of ${\rm VEL}[\delta(r)]$, a sphere with radius $r^{\star}$ in redshift space is approximately transformed into an oblate spheroid in real space with the smallest 
semiaxes, $r_{\parallel}$, along the line of sight. This semiaxes is related to $r^{\star}$ by:
 
\begin{equation}
r_{\parallel}=r^{\star}(1+{\rm VEL}[\delta(r)])^{-1}  \label{eq:eq17}
\end{equation} 
and
\begin{equation}
r_{\perp}=r^{\star}  \label{eq:eq18}
\end{equation} 
Now, in equation (\ref{eq:eq10}) the dependence on $r$ enters through $P(\delta_{l},r)$ and through $\delta(\delta_{l},r)$ (a small dependence) and in both of this quantities (see Appendix A) $r$ enters through $\sigma(Q)$ (the $rms$ of the linear enclosed density contrast within a Lagrangian sphere with radius $Q$). The relevant value of $Q$ here is $r[1+\delta(r)]^{1/3}$. Thus, in equation (\ref{eq:eq10}) the radius $r$ enters in the combination $\sigma(r[1+\delta(r)]^{1/3})$. In this equation, $r$ is the 
radius of a sphere in real space, but as we want the $P_{n}(r^{\star})$ on redshift space, the relevant body in real space, is no longer a sphere but an spheroid. 
The relevant $\sigma(Q)$ must now be evaluated within an spheroid. However, Betancort-Rijo \& Lopez-Corredoira (2002) showed that for spheroids not differing much 
from spheres $\sigma$ is almost independent of the form of the spheroid, depending only on its volume. Thus, we may use for $r$ the radius of a sphere with the same 
volume as the spheroid, which is given by

\begin{equation}  
r=(r_{\perp}^2r_{\parallel})^{1/3}=r^{\star}(1+{\rm VEL}[\delta(r)])^{-1/3}  \label{eq:eq19}
\end{equation}  

To obtain $P_n(r)$ in redsift space, that we represent by $P^{\star}_{n}(r^{\star})$, within this approximation, we may use equation (\ref{eq:eq10}) with the following replacement:  

\begin{equation} 
[1+\delta(\delta_{l},r)] \rightarrow [1+\delta(\delta_{l},r)][1+{\rm VEL}(\delta_{l})]  \label{eq:eq20} 
\end{equation}  
\\  
The reason for this replacement is that the procedure followed to obtain equation (\ref{eq:eq10}), which corresponds to real space can also be followed in redshift space. The 
only difference with real space is that in redshift space the role of the density contrast, $\delta$, is played by an apparent density contrast $\delta'$ 
(the right hand side term of equation (\ref{eq:eq20}) is simply $1+ \delta'$). 

The proposed replacement only takes into account the redshift distortion due to the smooth spherical 
velocity field associated with the underdensity within the sphere under consideration. The fluctuation of the velocity field around its mean is the source of an 
additional effect on the values of $P_{n}(r)$. We found using numerical simulations that the rms radial displacement of halos surrounding voids is $1.4 \mpch$ (Patiri et al. 2006). 
As a consequence of this random displacement some halos are driven away from the void, as defined in real space, while others are drawn closer to the void. So, 
since the density profile around a void is increasing, there are more halos being pulled in than out and the net effect is a contraction of the voids, or alternatively, 
a value of $P^{\star}_{n}(r^{\star}=r)$ smaller than $P_{n}(r)$. This effect is much smaller than the previous one, specially for the rather rare voids we are interested in. We suggest that this effect may be accounted for by a replacement in equation (\ref{eq:eq10}) of the form:

\begin{equation} 
\bar{n} \rightarrow \bar{n}\bigg[1+A\bigg(\frac{1.4 \mpch}{r}\bigg)^{2}\bigg]  \label{eq:eq21}
\end{equation}
where $A$ is a constant. We find, using numerical simulations, $A\simeq 6$. Using these replacements (equations  [\ref{eq:eq20}] and [\ref{eq:eq21}]) in equation (\ref{eq:eq10}) we find results for $P^{\star}_{n}(r^{\star})$ and the number density of voids that are in excellent agreement with the numerical simulations analyzed in this work (see Appendix A for details). Although these replacements represents the correct procedure, we shall also use, for computational reasons, an alternative procedure, in which we simply 
replace $r$ by certain function of $\delta$ (equation A9 in Appendix A), finding very similar results to those found using the procedure described above.

\section{Dependence of the number density of voids on redshift and cosmological parameters}

In this section we focus our attention on how the VPF and the number density of voids larger than a given radius depend on redshift and cosmological parameters $\sigma_{8}$ and $\Gamma$. Here we present the basics of the idea, referring the reader to Appendix A for details on the explicit procedure. In equation (\ref{eq:eq10}) we can see that the VPF depends on the cosmological parameters through the linear {\it rms}, of density fluctuations [$\sigma(r)$] which enters through the probability distribution for $\delta_{l}$ on the scale $r$.  There is also a dependence entering through $\delta_{ns}$ given in equation (\ref{eq:eq8}). 

The \LCDM~ transfer function (e.g. Bardeen et al. 1986) is essentially determined by a single parameter, $\Gamma$, which determine the co-moving horizon scale at matter domination. The barion density play a small role and is negligible to most of the cases in which our framework applies. The relevant parameters determining the power spectrum are $\sigma_{8}$ (its amplitude) and $\Gamma$ (its shape). In Appendix A we give $\sigma(r)$ as an explicit function of this two parameters, i.e. $\sigma(r,\sigma_{8},\Gamma)$. It is important to note that the coefficients $A(m)$ and $b(m)$ in equation (\ref{eq:eq8}) also depends on $\sigma_{8}$ and $\Gamma$ (see equation \ref{eq:eqA5}).

\begin{figure}
\includegraphics[width=\columnwidth]{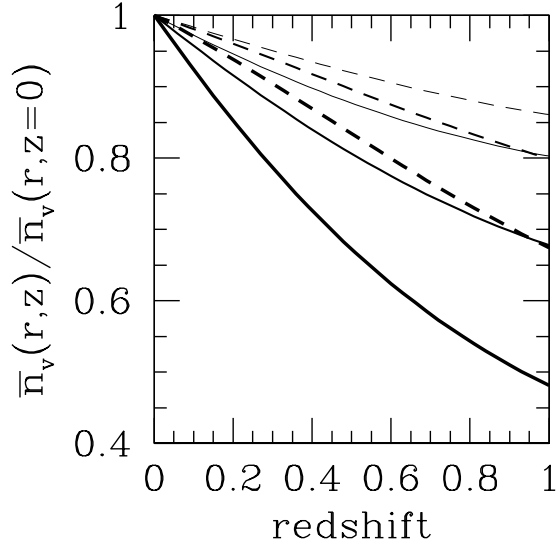}
\caption{The number density of voids larger than $11, 13$ and $15 \mpch$ (thinest to thickest lines respectively) defined by galaxies with number density $5 \times 10^{-3} (\mpch)^{-3}$ as a function of redshift. Full lines denote voids in real space, and dashed lines redshift space.}
\label{Gs}
\end{figure}

Once we have determined the dependence of the VPF and the number density of voids with the cosmological parameters, the redshift dependence for a given set of parameters may be obtained replacing $\sigma_{8}$ by:

\begin{equation}
\sigma_{8}\frac{D(z)}{D(z=0)} \quad ; \quad \bar{n}_{v}(r,z)=\bar{n}_{v}\left(r,\sigma_{8}\frac{D(z)}{D(z=0)}\right) \label{eq:eq36}
\end{equation}
where $D(z)$ is the linear growth factor of density fluctuations in the model under consideration. $\bar{n}_{v}(r,\sigma_8)$ is $\bar{n}_{v}(r)$ as a function of $\sigma_{8}$. This equation reflects the fact that the dependence of the void statistics (in real space) with redshift enters only through $\sigma_{8}$. 
However, the void statistics in redshift space shows a substantially smaller dependence on redshift (up to $z \simeq 1$).
We can see the difference with an example. In figure \ref{Gs}, we show the number density of voids larger than $11, 13$ and $15 \mpch$ (thinest to thickest lines respectively) defined by galaxies with number density $5 \times 10^{-3} (\mpch)^{-3}$ as a function of redshift, both in real and redshift spaces (full, dashed lines respectively). 
Note that the sample of galaxies defining the voids is selected to keep their number density fixed at any redshift. Therefore, the value of $m_g$ changes as $m$ 
changes with redshift. Also, for a given value of $m$, $\sigma(m)$ changes with $z$ (see Appendix A).

\section{Corrections due to observational effects}

\subsection{Voids within a finite sample}

The quantity $\bar{n}_{v}(r)$ represents the mean number density of non-overlapping maximal spheres with radius larger than $r$. These maximal spheres can be determined precisely for an arbitrarily large sample but this is not the case for finite samples. Consider a large sample (much larger than the mean distance between the voids) and assume that all maximal spheres larger than $r$ are located sufficiently far away from its boundaries. Consider now a smaller sample entirely contained within the region containing these maximal spheres. The mean number of maximal spheres larger than $r$ that are entirely contained within the finite sample is given by:

\begin{equation} 
\bar{N}(r)=\int_{r}^{\infty} -AV(r')~ \left(\frac{dn_{V}(r')}{dr'} ~dr'\right)  \label{eq:eq21b}
\end{equation}  
where $AV(r)$ is the available volume within the sample for the centers of the spheres of radius $r$, and the parenthesis is the number of maximal spheres with radius between $r$ and $r+dr$. Integrating by parts the r.h.s. of  equation (19) we have:

\begin{eqnarray} 
\bar{N}(r)&=&n_{v}(r)AV(r)+\int_{r}^{\infty} n_{V}(r')~ \frac{dAV(r')}{dr'}~ dr' \nonumber\\
          &\simeq& n_{v}(r)AV(r)+ \frac{dAV(r)}{dr} \int_{r}^{\infty} n_{V}(r')~dr' .  \label{eq:eq22}
\end{eqnarray}  
\\
This last approximation is valid only for values of $r$ such that the mean size $\bar{r}'$ of all voids larger than $r$ is only slightly larger than $r$. 
For $\bar{r}'$ we have:

\begin{eqnarray} 
\bar{r}'(r)&=& \frac{1}{\bar{n}_{v}(r)}\int_{r}^{\infty} -r'~ \frac{d\bar{n}_{v}(r')}{dr'} ~dr' \nonumber\\ 
       &=& r + \frac{1}{\bar{n}_{v}(r)}\int_{r}^{\infty} \bar{n}_{v}(r') ~dr'   \label{eq:eq23}
\end{eqnarray}  
Using this in equation (\ref{eq:eq22}) we have:

\begin{equation} 
\bar{N}(r)\simeq \bar{n}_{v}(r)AV[\bar{r}'(r)].   \label{eq:eq24}
\end{equation}  
\\
For simplicity, here we will consider samples that are either a box of side $L$ or a wedge defined by two parallels and
two meridians in the sky, with depth $R$. For the first case $AV(r)$ is given by:

\begin{equation} 
AV(r)=(L-2r)^3   \label{eq:eq25}
\end{equation}  
while for the second case we have (see Patiri et al. 2006):

\begin{equation} 
AV(r)=\int_{\frac{r}{sin(\Delta\delta/2)}}^{R-r} P(u,r) u^2 ~du   \label{eq:eq26}
\end{equation}  
where
\begin{eqnarray} 
P(u,r)= \newline 
\bigg[sin\big(\delta_{0}+\Delta\delta-sin^{-1}(r/u)\big)- \nonumber\\
- sin\big(\delta_{0}+asin(r/u)\big)\bigg] \times \nonumber\\
\times \bigg[\Delta\alpha-2~asin\bigg(\frac{asin(r/u)}{cos(\delta+\Delta\delta/2)}\bigg)\bigg] . \label{eq:eq27}
\end{eqnarray}  
$\delta_{0},\delta_{0}+\Delta\delta$ are the limits of the sample in declination while $\alpha,\alpha+\Delta\alpha$ are the limits in right ascension. 
The radial limit is $R$.

Equation (\ref{eq:eq24}) gives approximately the mean number of maximal spheres larger than a given radius $r$ within the sample when those maximal spheres have been determined within a sample much larger than the actual one. In practice, however, the maximal spheres are determined using only the actual sample. Thus, the maximal spheres close to the boundaries may, actually, not be maximal spheres but {\it locally} maximal spheres. These spheres are slightly smaller than the  actual maximal spheres and their centers are biased towards the center of the sample with respect to the centers of the actual maximal spheres. To avoid this effect one could simply discard all the locally maximal sphere which are so close to the boundaries that it can not be known by certain that they are actually maximal spheres.
But in this manner much information is lost, specially in narrow wedge samples. So, the appropriate thing to do is to use all locally maximal spheres and account for the border effect. 

As we pointed out before, the net effect of the boundary is to reduce slightly the size of the spheres and to retract their centers. The first effect results in a smaller value of $n_{v}(r)$ for any value of $r$, but the second produce the opposite effect as it is enclosing the centers of the spheres in a smaller volume. This last effect dominates, so that the number density of locally maximal spheres within a finite sample is larger than that for the actual maximal spheres.

This boundary effect can be taken into account modifying the available volume by:

\begin{equation} 
AV_{L}(r)=AV(\bar{r}'(r)-\alpha[\bar{r}'(r)-r])   \label{eq:eq28}
\end{equation}
where $AV_{L}(r)$ represent the available volume for locally maximal spheres, and $\alpha$ is a number between 2 and 4. The number we obtain in the Millennium Run numerical simulation is 3.3. Thus, we have:

\begin{eqnarray}
AV_{L}(r)=AV(r-2.3[\bar{r}'(r)-r]) & r \geq r_0 \nonumber\\
AV_{L}(r)=AV(r_0-2.3[\bar{r}'(r_0)-r_0]) & r \leq r_0   \label{eq:eq29}
\end{eqnarray}
where $r_0$ is defined by:
\begin{eqnarray}
K(r_0)=0.34
\end{eqnarray}
and $K$ is defined in equation (4). The interesting cases are those with $r\geq r_0$, but written in this form the expression is valid even for very common voids, at least down to the largest $K$ values we explored ($K=0.43$). Finally, we have for the mean number of locally maximal spheres within the sample, $\bar{N}_{L}(r)$,

\begin{equation} 
\bar{N}_{L}(r)= \bar{n}_{v}(r)AV_{L}(r)   \label{eq:eq30}
\end{equation}
where $\bar{n}_{v}(r)$ is the mean number density of actual maximal spheres (given by equation 3) and $AV_{L}(r)$ is given by (\ref{eq:eq29}). We shall use this expression to estimate the sample independent quantity, $\bar{n}_{v}(r)$, from the observed statistics, ${N}_{L}(r)$. Note that the actual number density of locally maximal spheres
within the sample, $\bar{n}_{L}(r)$, is given by:

\begin{equation} 
\bar{N}_{L}(r)= \bar{n}_{L}(r)AV[\bar{r}'(r)]  \label{eq:eq31}
\end{equation}
but $\bar{n}_{L}(r)$ is a sample dependent quantity. 

In what follows we drop the distinction between local and actual maximal spheres. Samples are always finite, so it is understood that some of the 'maximal spheres' found in them, those close to the border, are only conditionally maximal. However, we use them to estimate the sample independent number density of actual maximal spheres.

\subsection{Spectroscopic completeness}

In the equations considered so far, the local number density of galaxies is modulated by the underlying density fluctuations and their corresponding bias. However, in real galaxy redshift surveys, there exist fluctuations in the local number density of galaxies due to the failure of taking spectra of some galaxies. The completeness,
defined as the ratio of successfully obtained redshifts to targetable objects, varies nontrivially from 0 to 1, having angular and magnitude dependencies. In this paper we assume a constant completeness in magnitude, to focus in the spectroscopic completeness. This issue have different origins and usually lay around 10\% depending on the survey (see e.g. Norberg et al. 2002; Conroy et al. 2005). These fluctuations, must be incorporated into our framework to obtain predictions for the number of voids expected within real galaxy samples. The essential question is whether the correlation for the completeness fluctuations is much larger than the size of the voids. If the completeness changes over the size of a void, the formalism developed in this work need to be re-elaborated to some extend. However, if the correlation length of the completeness fluctuations is much larger than the size of the 
voids, these fluctuations behave as a random dilution of the number of galaxies. Thus, Equation (10) can be used replacing $\bar{n}$ by its diluted value. So, $P_{n}(r)$ 
is given by:

\begin{equation}
P_{n}(r)=\int_{0}^{1}[P_{n}(r,\bar{n}c)]~P(c)~dc, \label{eq:eq32}
\end{equation}  
where the function $P_{n}(r,\bar{n})$ is the same as in equation (10) and $P(c)$ is the probability distribution for the spectroscopic completeness, $c$. $\bar{n}$
denotes the intrinsic mean number density of galaxies. Note that the observed density within the sample is $\bar{n}c$. To obtain the number density of
maximal spheres, we have to change $P_{n}(r)$ by $\bar{n}_{v}$ in the previous equation.

\subsection{The {\it snapshot} effect}

The equations presented above compute the predictions for the number of voids assuming a constant value of $z$ over the sample. However, in observational samples there is a dependence of redshift with the radial coordinate. Thus, due to the dependence of voids number densities with redshift, there exist a radial dependence on those densities.

The number density of voids at a distance $u$ from the origin is given by
$\bar{n}_{v}[r,z(u)]$ where $z(u)$ is the relationship distance-redshift. For a rectangular strip the expected number of voids is given by:

\begin{equation}
\bar{N}_{V}(r)=\int_{\frac{\bar{r}}{sin(\Delta\delta/2)}}^{R-\bar{r}}  \label{eq:eq33}
P(u,r) \bar{n}_{v}[r,z(u)] u^2 du
\end{equation}  
where
\begin{equation}
\bar{r}=r-2.3[\bar{r}'(r)-r] \label{eq:eq34}
\end{equation}
where $P(u,r)$ is given in (\ref{eq:eq27}); $\bar{r}'$ is given in (\ref{eq:eq23}), and $R$ is the depth of the sample. For small values of $z$, $\Omega_{m}=0.3$, $\Omega_{\lambda}=0.7$, $z(u)$ may be approximated by:

\begin{equation}
z(u) \simeq 3.336\times 10^{-4} u + 2.7\times 10^{-8} u^2  \label{eq:eq35}
\end{equation}
with co-moving distance $u$ in units of $\mpch$. The mean number density of voids within the sample is obtained dividing the expected number of voids by the volume available for those voids, $AV_{L}(r)$, given by equation (\ref{eq:eq29}).

\section{Tests to estimate cosmological parameters}

The framework presented here can be used to perform tests to measure the cosmological parameters. The efficiency of any test can be addressed rigorously using our formalism, so that we may design an optimum test. Here, we will consider tests built using the VPF and the number density of voids. 

\subsection{VPF {\it vs.} number density of voids}

For both the VPF and the number density of voids larger than a given radius, most of the information concerning cosmological parameters comes from extreme events. 
However, implementing all the useful information (i.e. including more common voids) can be used to improve the statistics. In order to determine the efficiency 
of a test to measure a given generic cosmological parameter, let say $\alpha$, we can construct 
a simple test considering only the voids larger than a given 
radius $r$. The $rms$ of that parameter within this test would be:

\begin{equation}
\frac{rms(\alpha)}{\alpha}=\frac{rms[\bar{n}_{v}(r)]}{\bar{n}_{v}(r)G(r,\alpha)} \label{eq:eq37}
\end{equation}
where
\begin{equation}
G(r,\alpha)=\frac{d\,{\rm ln}\,\bar{n}_{v}(r)}{d\,{\rm ln}\,\alpha} \label{eq:eq38}
\end{equation}

The fractional $rms$ error for the estimate of $\bar{n}_{v}(r)$ for a given sample is that corresponding to an uniform Poissonian distribution with a correction term, assuming that voids are uncorrelated (which seems to be the case, see Patiri et al. 2006). Therefore, 

\begin{equation}
\frac{rms[\bar{n}_{v}(r)]}{\bar{n}_{v}(r)}=\frac{1}{\bar{N}(r)^{1/2}}[1-3[\bar{r}'(r)]^3 \bar{n}_{v}(r)]^{1/2} \label{eq:eq39}
\end{equation}
where $\bar{N}(r)$ is the expected number of voids larger than a given radius $r$. The last parenthesis accounts for the anticorrelation for voids with distances between their centers less than $2\bar{r}'$. This factor is important only for small voids. Then, the best test using this void statistic (i.e. all voids larger than $r$) is obtained by minimizing equation (\ref{eq:eq37}) with respect to $r$. For the relevant range of parameters, the same value of $r$ is obtained for any value of $\alpha$. The efficiency of this 
test is somewhat smaller than that of the best test using all the voids within the sample, but it remains as simple as for a first approximation.

In a similar way, to study the ability of the VPF to estimate the parameter $\alpha$ we may consider a simple test using the value of the VPF for a given value of $r$. 
In this case, we have for the error:

\begin{equation}
rms(\alpha)=\frac{rms[P_{0}(r)]}{P_{0}(r)G'(r,\alpha)} \label{eq:eq40}
\end{equation}
where
\begin{equation}
G'(r,\alpha)=\frac{d\,{\rm ln}\,P_{0}(r)}{d\,{\rm ln}\,\alpha}  \label{eq:eq41}
\end{equation}
In Patiri et al. (2006) we have shown that

\begin{equation}
rms[P_{0}(r)]=2.82\frac{rms[\bar{n}_{v}(r)]}{\bar{n}_{v}(r)}P_{0}(r).  \label{eq:eq42}
\end{equation}
We then have for the $rms(\alpha)$:
\begin{equation}
rms(\alpha)=2.82\frac{rms[\bar{n}_{v}(r)]}{\bar{n}_{v}(r)}  \label{eq:eq43}
\end{equation}
Example values of $G$, $G'$ are given in table \ref{table:tbGs}.

\subsection{Maximum Likelihood test}

Another test is the Maximum Likelihood test. To perform this test, we consider all voids larger than a given radius. This radius could be chosen arbitrarily small, 
taking in account that very small voids do not carry any cosmological information. The probability distribution for the radius of all voids larger than $r_{0}$ is 
given by:

\begin{equation}
P(r)=-\frac{1}{\bar{N}(r)}\frac{d \bar{N}(r)}{dr}  \label{eq:eq44}
\end{equation}
where $\bar{N}(r)$ is the expected value for the number of voids, larger than $r$, within the sample (for a given set of cosmological parameters).

As in this paper we are particularly interested in $\sigma_{8}$ and $\Gamma$, the likelihood of 
the observed data for that set of parameters is

\begin{equation}
L(\sigma_{8},\Gamma)=\prod_{i=1}^{n} P(r_{i},\sigma_{8},\Gamma).  \label{eq:eq45}
\end{equation}
This is so because the sizes of two different voids are independent random variables. With this function the best estimate may be obtained maximizing with 
respect to $\sigma_{8}$ and $\Gamma$ while the confidence levels are obtained following the standard Bayesian approach.

\subsection{$\chi^2$ test}

A less sophisticated test, but almost as efficient as that of Maximum Likelihood, is the $\chi^{2}$ test. 
This test can be done computing the $\chi^{2}$, where the number of degrees of freedom corresponds to the number of bins in radius:

\begin{equation}
\chi^{2}(\sigma_{8},\Gamma)=\sum_{i=1}^{n} \frac{[N_{i}-\bar{N}_{i}(\sigma_{8},\Gamma)]^{2}}{[rms(N_{i})]^{2}},  \label{eq:eq46}
\end{equation}
where $N_{i}$ is the number of voids with radius within the $i$-th bin found in the observational sample. On the other hand, the expected value for the number of voids in 
the same bin in radius for given values of $\sigma_{8},\Gamma$ is:

\begin{eqnarray}
\bar{N}_{i}(\sigma_{8},\Gamma) = \bar{N}_{V}(r_{i+1})(\sigma_{8},\Gamma)-\nonumber\\
-\bar{N}_{V}(r_{i})(\sigma_{8},\Gamma)  \label{eq:eq47}
\end{eqnarray}
where $r_{i+1}$ and $r_{i}$ are the boundaries of the $i$-th bin and $\bar{N}_{V}(r)$ is given in equation (\ref{eq:eq31}). The $rms$ of $N_{i}$ is given by equation (\ref{eq:eq39}) (note the change in the subscript $L$ by $V$) and it is well approximated by $\bar{N}_{i}$.

The statistical significance of the $\chi^{2}$ test is that of a $\chi^{2}$ statistic with $n$ degrees of freedom because the number of voids in each bin is an independent 
random variable. The width of the bins must be chosen wide enough to contain a large number of voids, so that this number may be assumed to follow a Gaussian distribution. Even though $\bar{N}_{V}(r)$ can be computed directly with our analytic equations, its accuracy is in the order of 5-7\% which could slightly change the result of the test.
To avoid this, we suggest to use mock catalogs with large volumes in order to better quantify the sampling errors. We can then use our analytic framework to {\it extrapolate} the statistics found in mock catalogs to other values of cosmological parameters. The values for $\bar{N}_{i}$ can be computed in the 
following way:
\begin{equation}
\bar{N}_{i}(\sigma_{8},\Gamma)=\bar{N}^{sim}_{i}{\rm F}(\sigma_{8},\Gamma) \label{eq:eq48}
\end{equation}
and
\begin{equation}
F(\sigma_{8},\Gamma)=\frac{\bar{N}_{i}(\sigma_{8},\Gamma)}{\bar{N}_{i}[\sigma_{8}^{sim},\Gamma^{sim}]}  \label{eq:eq49}
\end{equation}
where $\sigma_{8}^{sim}$, $\Gamma^{sim}$ are the cosmological parameters for which the numerical simulation was ran. Note that ${\rm F}$ is simply the ratio between the expected values obtained with our analytical 
expression for the values of the corresponding cosmological parameters and those values obtained in 
the numerical simulation.

\section{Comparison with Numerical Simulations}
\label{res}

We have to use cosmological numerical simulations extensively in order to test the predictions 
made by our formalism. We also need these simulations to calibrate the equations with free parameters. 

\subsection{Cosmological $N$-body Simulation and Galaxy Formation Model}

In this work we took advantage of the publicly available dark matter halo and mock galaxy catalogs produced with the Millennium Run simulation (Springel et al. 2005) and the `MPA' semi-analytic model of galaxy formation (SAM) applied to it (Croton et al. 2006; De Lucia \& Blaizot 2007). The Millennium Run simulation follows the evolution of $10^{10}$ dark matter particles in a periodic box of $500 \mpch$ on a side with a mass resolution per particle of $8.6 \times 10^8 \Msunh$. The initial conditions of the simulation were run with cosmological parametes consistent with the combined analysis of the 2dFGRS and WMAP1 data ($\Gamma=0.1825$, $\sigma_8=0.9$). The halos in the simulation were identified in each time step using a friend-of-friends algorithm with a linking length of 0.2 times the mean particle separation. For full details of the simulation we refer the reader to Springel et al. (2005).

The gravitational potential of each halo accretes the surrounding gas from which galaxies form. The semi-analytic models tracks, in a parametrized way, various physical processes that are supposed to play a key role in galaxy formation such as radiative cooling of hot gas, star formation in the cold disk, supernova feedback, black hole growth and AGN feedback through the `quasar' and `radio' epochs of AGN evolution, metal enrichment of the inter-galactic and intra-cluster medium, and galaxy morphology shaped through mergers and merger-induced starbursts.

The full galaxy catalog produced with the SAM (also known as `Delucia2006a' catalog\footnote{Both dark matter and galaxy catalogs can be downloaded from http://www.g-vo.org/MyMillennium2/}) contains information for about 12 million galaxies brighter than $M_r=-17$. For each of these galaxies we have available, among other properties, positions and velocities, magnitudes in several band passes (Johnson, Busher, 2MASS as well as the 5 SDSS bands), stellar mass and the mass of its parent dark matter halo. 

In order to quantify errors, we divided the full $500 \mpch$ on a side box in 8 small boxes of $250 \mpch$ side each. From these boxes we constructed halo and galaxy samples, in both real and redshift space. To produce redshift space catalogs we used the {\it distant observer} technique. We selected for our samples all halos with masses larger than $6.6 \times 10^{11} \Msunh$ and galaxies brighter than $M_{r} = -20.4 +5logh$. The mean number density of objects in our halo and galaxy samples is $5$ and $5.17 \times 10^{-3} \mpch$ respectively.

\subsection{Results}

We conducted extensive comparisons for $P_n(r)$ and  $\bar{n}_v(r)$ in the galaxy and halo distributions, in both real and redshift spaces, finding excellent agreement. 

\begin{figure*}
\includegraphics[width=165mm]{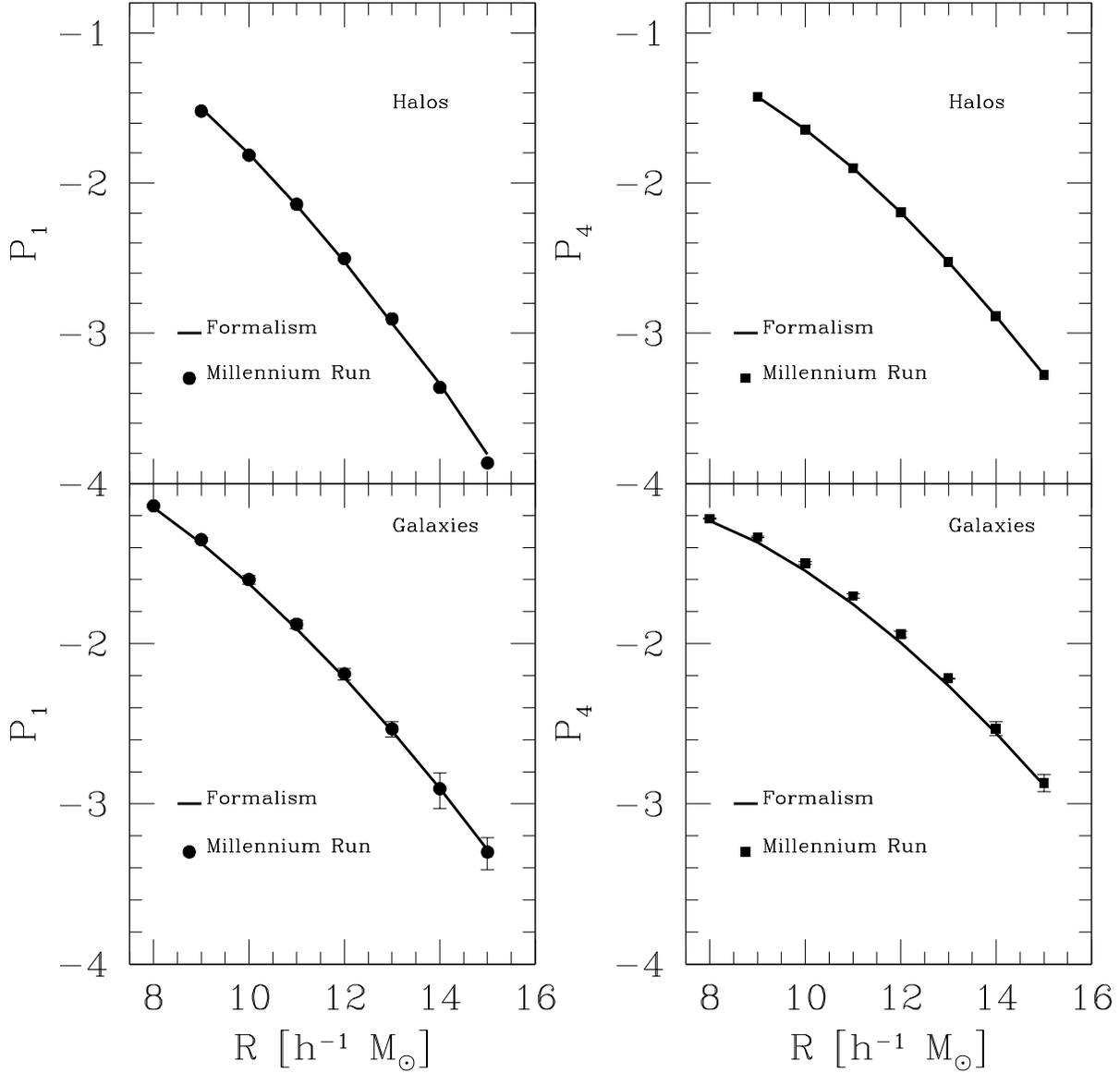}
\caption{Top panels: $P_1(r)$ and $P_4(r)$ in the distribution of halos with mass larger than $6.6 \times 10^{11} \Msunh$ in {\it real space}. The full line and symbols are the results obtained using our formalism and the Millennium Run respectively. The number density of halos in this sample is $5\times 10^{-3} \mpch$. Bottom panels: $P_1(r)$ and $P_4(r)$ obtained with our formalism and in the mock catalog for galaxies brighter than $M_{r} = -20.4 +5log h$ in {\it redshift space}. The symbols are the same than in the top panels.}
\label{fig:fig1}
\end{figure*}

In Figure \ref{fig:fig1} (top panels) we show $P_1(r)$ and $P_4(r)$ as a function of radius in the distribution of dark matter halos with mass larger than $6.6 \times 10^{11} \Msunh$ in {\it real space}. The full line and symbols are the results obtained using our formalism and the Millennium Run respectively. The number density of halos in this sample is $5\times 10^{-3} \mpch$. In the bottom panels we show $P_1(r)$ and $P_4(r)$ obtained for the galaxies in {\it redshift space}. The symbols are the same than in the top panels. The number density of galaxies is $5.17 \times 10^{-3} \mpch$ and the mass of the halos that are clustered as these galaxies is $m_g=9.14 \times 10^{11} \Msunh$ [see eq. (\ref{eq:eqA6})]. 
We can see that, overall, the results obtained with our formalism are in excellent agreement with the statistics found in the Millennium Run both for halos and galaxy distributions.  

\begin{figure}
\includegraphics[width=\columnwidth]{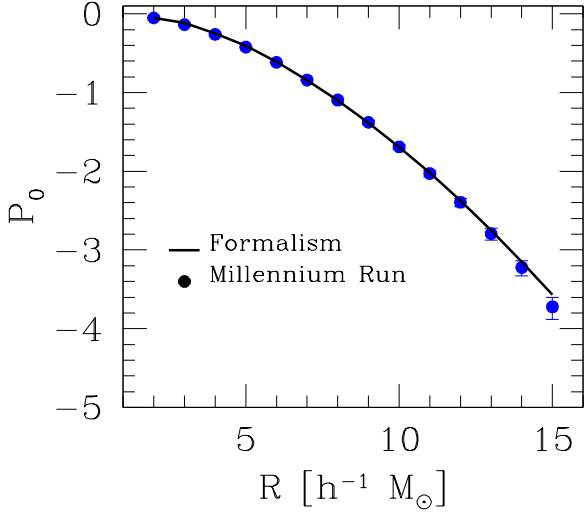}
\caption{The void probability function (VPF) for the same galaxies as figure (\ref{fig:fig1}). See text for details.}
\label{fig:fig2}
\end{figure}

In Figure \ref{fig:fig2} we show the VPF for the same distribution of galaxies mentioned above. We can see that our formalism is reliable even to values of $r$ much smaller than those actually relevant for our purposes. Also, the fact that galaxies are clustered like halos above some mass (see eq. [\ref{eq:eq14}]), not only produce the correct VPF, but also leads to the correct value for other statistics of underdense regions on the relevant scales, i.e. those larger than the halos themselves. 

\begin{figure*}
\includegraphics[width=165mm]{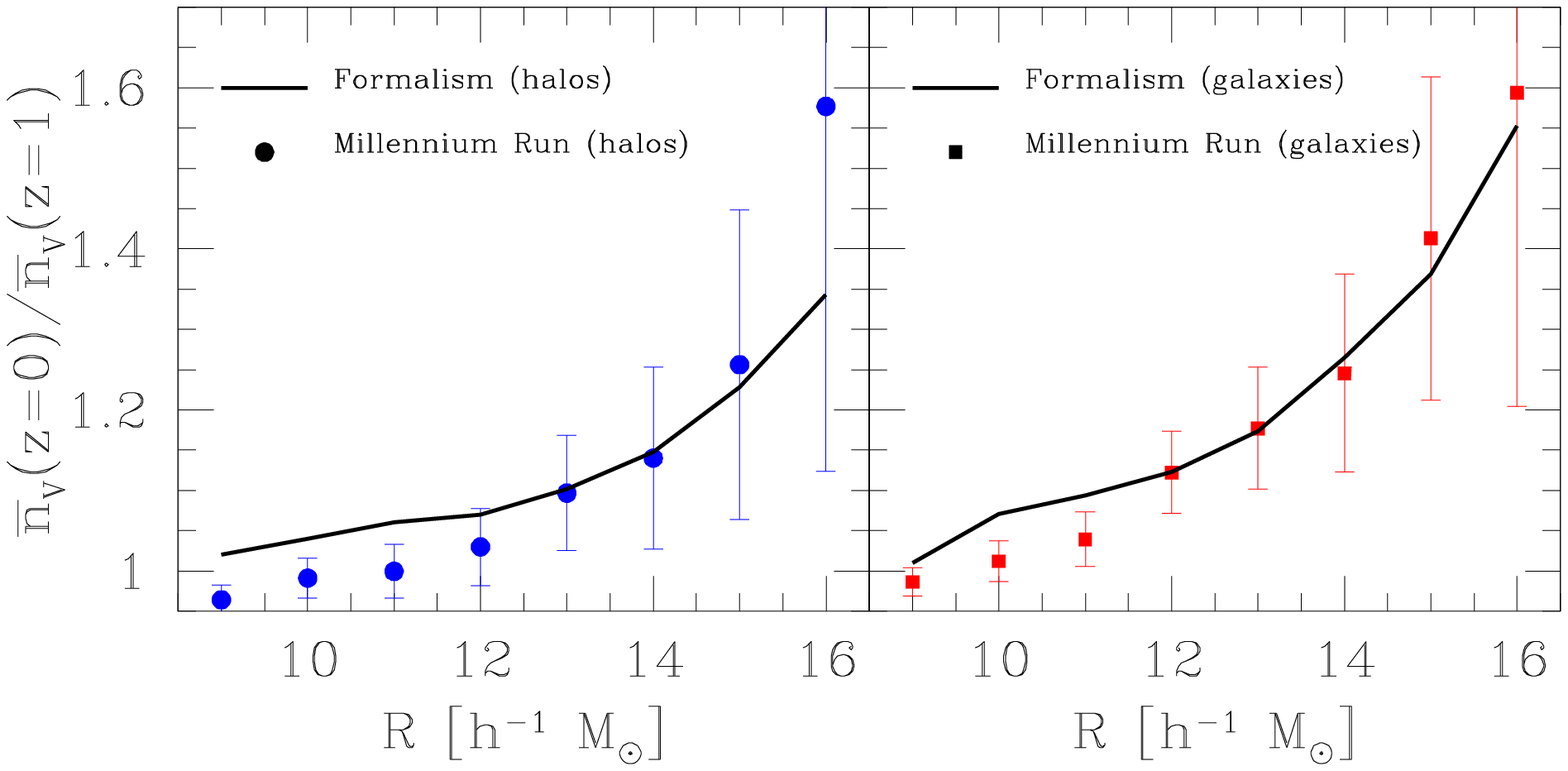}
\caption{Left panel: The ratio of the number density of voids at present to those at $z=1$ in the distribution of dark matter halos obtained using our formalism (full line)  and the Millennium Run (discrete symbols). Right panel: the same ratio but for voids defined by galaxies. In both cases the value of the number density of the defining objects is $5 \times 10^{-3} \mpch$.}
\label{fig:fig3}
\end{figure*}

In Figure \ref{fig:fig3} we show the comparison of predictions for the number densities of voids larger than $r$ obtained from our formalism and those from the Millennium Run in redshift space. In the left panel of this figure we show the ratio of the number density of voids at present to those at $z=1$ in the distribution of dark matter halos. In the right panel we plot the same ratio but for voids defined by galaxies. In both cases the value of the number density of the defining objects is $5 \times 10^{-3} \mpch$. At $z=1$, the mass (lower limit) of the halos with the same number density as those at $z=0$ is $m=6.48 \times 10^{11} \Msunh$ (see eq. A9). 
For galaxies, we have [using eq. (\ref{eq:eq14})] $m_g=7.93 \times 10^{11} \Msunh$. We see that $\bar{n}_v$ is accurately predicted by our formalism both for voids defined by halos and galaxies. 
Also, the dependence of redshift is obtained with good accuracy for all voids, specially for those rare enough [$K \la 0.2$, see eq. (\ref{eq:eq4})] and relevant for 
our purposes.

It is important to note that the predictions for $P_n(r)$ in halo distributions obtained using our formalism do not involve any fit to the simulations. However, in order to compute $\bar{n}_v(r)$ using equation (\ref{eq:eq3}), we have to fit two coefficients (3.5 and 2.18 in that equation).  This was done using the Millennium Run and several other results from numerical simulation (see Patiri et al. 2006). These coefficients are particularly important for common voids (e.g. voids larger than $12 \mpch$ defined by the galaxies used above). However, for rare voids, the relevance of these coefficients is much smaller.

For galaxy distributions, our predictions for $P_n(r)$ and $\bar{n}_v(r)$ involves determining the mass of the halos ($m_g$) which are clustered like the galaxies. To this end we use $m_g$ as a free parameter and choose it so as to maximize the agreement between the predictions for $P_n(r)$ and $\bar{n}_v(r)$ using our formalism and the results found in the numerical simulations. We obtain $m_g$=$9.15 \times 10^{11} \Msunh$, and with this value we fitted the only free parameter in equation (\ref{eq:eq14}). Using this equation we may obtain the value of $m_g$ corresponding to any galaxy sample at any redshift and for any value of the cosmological parameters.

Finally, in Table \ref{table:tbGs} we show the quantities ${\rm G}_{\Gamma}$, ${\rm G}'_{\Gamma}$, ${\rm G}_{\sigma_{8}}$, ${\rm G}'_{\sigma_{8}}$, ${\rm G}_{\bar{n}}$ and  ${\rm G}'_{\bar{n}}$  generically defined by:
\begin{equation}
G_{\alpha} \equiv G(r,\alpha) ; \quad G'_{\alpha} \equiv G'(r,\alpha)  \label{eq:eq50}
\end{equation}
with $G(r,\alpha)$, $G'(r,\alpha)$ as defined in equations (\ref{eq:eq38}) and (\ref{eq:eq41}). These quantities characterize the sensitivity of $\bar{n}_v(r)$ and $P_0(r)$ to the generic parameter $\alpha$. We also consider the quantities $G_{\gamma}$ and $G'_{\gamma}$:\\

\begin{eqnarray}
G_{\gamma} \equiv G(r,\gamma) = \frac{d {\rm ln}\bar{n}_v(r)}{d {\rm ln}\gamma} \nonumber\\
G'_{\gamma} \equiv G'(r,\gamma)= \frac{d {\rm ln}P_0(r)}{d {\rm ln}\gamma} \nonumber \\
\gamma \equiv \frac{d {\rm ln}\sigma(r)}{d {\rm ln}r}     \label{eq:eq51}
\end{eqnarray}
These quantities are closely related to $G_{\Gamma}$, $G'_{\Gamma}$ and characterize the sensitivity of $\bar{n}_v(r)$, $P_0(r)$ to the local logarithmic slope of $\sigma(r)$. They may be obtained from $G_{\Gamma}$, $G'_{\Gamma}$ through the relationship between $\Gamma$ and $\gamma(r)$.

\begin{table*}
\begin{tabular}{|c|rrrrrr|}
\hline
Radius $[\mpch]$  & ${\rm G}_{\Gamma}$ & ${\rm G}'_{\Gamma}$ &  ${\rm G}_{\sigma_{8}}$   &  ${\rm G}'_{\sigma_{8}}$ & ${\rm G}_{\bar{n}}$ & ${\rm G}'_{\bar{n}}$ \\
\hline

  9.0 &         &  0.726 &  0.047 & 1.194 & 0.794 & 1.062  \\
 10.0 &         &  1.030 &  0.406 & 1.459 & 0.580 & 1.203  \\   
 11.0 &         &  1.393 &  0.594 & 1.748 & 0.829 & 1.346  \\   
 12.0 &   0.149 &  1.816 &  0.718 & 2.065 & 0.810 & 1.500  \\   
 13.0 &   0.430 &  2.304 &  0.904 & 2.412 & 0.898 & 1.661  \\    
 14.0 &   0.814 &  2.857 &  1.207 & 2.790 & 1.021 & 1.825  \\    
 15.0 &   1.327 &  3.480 &  1.438 & 3.201 & 1.318 & 1.996  \\    
 16.0 &   1.953 &  4.174 &  2.086 & 3.646 & 1.563 & 2.170  \\

\hline
\end{tabular}
\caption{The values of $G$ and $G'$ represent the sensitivity of the number densities of voids and the VPF to the parameters shown respectively. 
These numbers correspond to $\sigma_{8}=0.9,\Gamma=0.1825$ and $\bar{n}=5\times 10^{-3} (\mpch)^{-3}$. See text for details.}
\label{table:tbGs}
\end{table*}

\section{Discussion and Conclusions}

In this work we developed an analytical formalism dealing with the clustering properties of dark matter halos and galaxies in underdense regions. 
In particular, we extended an existing framework to account for redshift distortions and observational effects. We also included in the formalism the high precision conditional mass function recently published by Rubi\~no et al. (2008). We showed that our formalism allows us to calculate accurately several void and underdense statistics, such as the $P_{n}(r)$ and $\bar{n}_{v}(r)$ in dark matter halo distributions in both real and redshift spaces at any redshift. Our predictions for $P_{n}(r)$ are particularly remarkable, since they are the result of purely theoretical considerations. 

We also found that for galaxy distributions, $P_{n}(r)$ and $\bar{n}_{v}(r)$ may be obtained, to a very good approximation, assuming that the galaxies have the same clustering properties as halos above a given mass $m_{g}$.
We deduced a relationship between this mass and that of halos with the same accumulated number density as the galaxies. Similar approaches have shown very useful to describe other properties of galaxy clustering (e.g. Conroy, Wechsler \& Kravtsov 2006).  
The equation we obtained contains a single free parameter that we fitted using numerical simulations at $z=0$, leading to predictions for $P_{n}(r)$ and $\bar{n}_{v}(r)$ that are remarkably good at any redshift compared to those statistics found in the simulations.

We found that the dependence of $P_{n}(r)$ and  $\bar{n}_{v}(r)$ on redshift is small, with $\bar{n}_{v}(r)$ changing less than 20\% between $z=1$ and $z=0$ for voids with radius larger than $13 \mpch$. This is due to the fact that in cosmologies with $\Omega_{m}<1$, the redshift distorsions are more effective (for a given value of $\delta$) at higher redshift. This partially compensates (up to $z \sim 1$) for the smaller amplitude of density fluctuations. However, the dependence of $P_{n}(r)$ and $\bar{n}_{v}(r)$ on $\sigma_{8}$ and $\Gamma$ is considerably larger, making them important to use as tests to measure these parameters. We showed how to construct efficiently several of these tests and discussed in detail the treatment of several observational effects. Correcting for the biases implied by these effects may be necessary for an accurate measurement of cosmological parameters by means of $P_{n}(r)$ and $\bar{n}_{v}(r)$.

From the cosmological parameter estimation point of view, there is a close symmetry between voids and clusters. In fact, rich clusters and large voids (which set the strongest constrains to cosmological parameters) correspond to regions of co-moving scale of $\sim 5 \mpch$. The main difference between them is that clusters collapse to form structures of about $\sim 2-3 \mpch$ while proto-voids expands to voids of $\sim 13 \mpch$. In both cases the efficiency lay on how precisely the mass of the underlying dark matter can be determined. For clusters, the main source of uncertainty comes through the different methods to determine their masses, such as the temperature-mass relation when using x-rays (Vikhlinin et al. 2008), while voids do not present this problem. Moreover, voids might be specially interesting to measure the normalization of the amplitude of density fluctuations since they are tracing comparable scales, while clusters are considerably smaller.

The currently available (2dFGRS, SDSS) and next generation (e.g. BOSS) large galaxy redshift surveys in combination with the analytical formalism presented in this paper, 
will allow us to estimate the values of the relevant cosmological parameters using void statistics, providing independent measurements. These, along with the estimations made by other methods will contribute to reduce their uncertainties. It is also worth to mention that in a forthcoming paper we will explore the dependence of the statistics of voids with $\sigma_{8}$ and $\Gamma$, using the void statistics found in the 2dFGRS (Patiri et al., in prep.). All this makes the statistics of underdense regions a very promising tool to constrain cosmology not just with the current surveys but also to next generation high redshift surveys.

\section*{Acknowledgments}

JBR, SGP \& FP thank support from the Spanish MEC under grant PNAYA 2005-07789. 

The Millennium Run simulation used in this paper was carried out by the
Virgo Supercomputing Consortium at the Computing Center of the Max-Planck
Society in Garching.

\appendix{}
\section{Explicit computation of $P_{\lowercase{n}}(\lowercase{r})$}

In this Appendix we show the detailed procedure for computing $P_n(r)$ and the void statistics for given values of $\sigma_8$ and $\Gamma$.

\begin{equation}
P_n(r) = \frac{1}{n!} \int_{-7}^{0} P(\delta_l,r) [u(\delta_l)]^n e^{[-u(\delta_l)]} d\delta_l   \label{eq:eqA1}
\end{equation}

\begin{equation}
u(\delta_l) = \bar{n} V [1+{\rm DELF}(\delta_l,r)] A e^{-b\delta_l^2}    \label{eq:eqA2}
\end{equation}
${\rm DELF}$ is a function of $\delta_l$ and $r$ that gives the mean actual density contrast within a sphere with radius $r$ with enclosed 
linear density contrast $\delta_l$.

\begin{eqnarray}
{\rm DELF}(\delta_l,r) \equiv & \nonumber\\
 \frac{1+{\rm DELT}(\delta_l)}{| 1 - \frac{4}{21}[1+{\rm DELT}(\delta_l)]^{2/3}[\sigma(r[1+{\rm DELT}(\delta_l)]^{1/3})]^{2} |} &    \label{eq:eqA3}
\end{eqnarray}
where ${\rm DELT}(\delta_l)$ denotes the relationship between the actual and linear enclosed density contrasts in the spherical colapse model 
(see PBP06 for details).

\begin{equation}
1+{\rm DELT}(\delta_l) \simeq (1-0.607\delta_l)^{-1.66}   \label{eq:eqA4}
\end{equation}
$\sigma(Q)$ is the rms of the linear density contrast on a sphere with Lagrangian radius $Q$. In this equation, $\sigma(Q)$ is evaluated 
at $Q$ equal to $r[1+{\rm DELT}(\delta_l)]^{1/3}$. Explicit equations for $\sigma$ are given at the end of this Appendix.
$A$, $b$ in eq. (A2) are also functions of $\delta_l$ given by:

\begin{eqnarray}
 A \equiv &A(m,Q=r[1+{\rm DELT}(\delta_l)]^{1/3})  \times  \\
 &\times \left(\frac{D(z)\sigma_8}{0.9}\right)^{0.88} \left(\frac{\Gamma}{0.21}\right)^{0.343} \nonumber \\
 \nonumber\\
 b \equiv &b(m,Q=r[1+{\rm DELT}(\delta_l)]^{1/3})  \times   \\
 & \times \left(\frac{D(z)\sigma_8}{0.9}\right)^{-2.55} \left(\frac{\Gamma}{0.21}\right)^{-0.82} \nonumber \label{eq:eqA5}
\end{eqnarray} 
\\
where $A(m,Q)$, $b(m,Q)$ are functions of the mass of the objects and the Lagrangian radius of the regions being considered (that 
corresponding to an Eulerian sphere with radius $r$, see Rubi\~no et al. 2008):

\begin{eqnarray}
A(m,Q) = &[1.577 - 0.298 (\frac{Q}{8})] - \\
- &[0.0557 + 0.0447 (\frac{Q}{8})] {\rm ln}(m) -\nonumber\\
- &[0.00565 + 0.0018(\frac{Q}{8}][{\rm ln}(m)]^2\nonumber\\
\nonumber\\
b(m,Q) = &[0.0025 - 0.00146 (\frac{Q}{8})] + \\
+ &[0.121 - 0.0156(\frac{Q}{8})] \times \nonumber \\
\times & m^{[0.335 + 0.019 (\frac{Q}{8})]} \nonumber\\
\nonumber\\
m \equiv &\frac{M}{3.51 \times 10^{11} h^{-1}M_{\odot}}\left(\frac{0.21}{\Gamma}\right)    \label{eq:eqA6}
\end{eqnarray}
where $M$ is the mass of the objects, $D(z)$ in equations A5,A6 is the linear growth factor normalized to be 1 at present. The exponents determining the dependence on $A$, $b$ on $\sigma_8$ and redshift are slightly different from those given by Rubi\~no et al. (2008) but are within the precision afforded by the procedure used in that work. The exponents given here have been accurately fitted using numerical simulations. 

For dark matter halos, the values of $m$ entering in these last equations is defined by:
\begin{equation}
\bar{n}(>m)=\bar{n}_{sample}
\end{equation}
while for galaxies we have to use $m_g$ given in equation (\ref{eq:eq14}). The definition of $m$ and $m_g$ imply that these 
quantities have to be scaled with $\sigma_{8}$, $\Gamma$ and redshift. However, $m$ changes very little with these variables up to $z\sim 1$, so we 
chose to held it fixed. In the case of $m_g$ the scaling can be approximated by:

\begin{equation}
m_{g}(\sigma_{8},\Gamma,z) = m (1.396)^{(D(z)\sigma_{8}/0.9)(\Gamma/0.21)^{0.271})}
\end{equation}

We use for the probability distribution $P(\delta_l,r)$ of the linear density contrast within an Eulerian space given by Betancort-Rijo \& L\'opez-Corredoira (2002):

\begin{eqnarray}
P(\delta_l,r) = &\frac{exp\left[\frac{-1}{2} \frac{\delta_l^2}{(\sigma(r[1+{\rm DELF}(\delta_l,r)]^{1/3})^2}\right]}{\sqrt{2\pi}} \times \nonumber\\ 
\times &[1 + {\rm DELF}(\delta_l,r)]^{-(1-\frac{\alpha}{2})} \times \nonumber\\
\times &\frac{d}{d\delta_l} \left( \frac{\delta_l}{\sigma(r[1+{\rm DELF}(\delta_l,r)]^{1/3})}\right)\nonumber\\
\alpha(\delta_l,r) = &0.54 + 0.173  \times  ln \left( \frac{r[1+{\rm DELT}(\delta_l)]^{1/3}}{10} \right)    \label{eq:eqA7}
\end{eqnarray}

Even though $\alpha$ depends on $\Gamma$, and the above equation corresponds to $\Gamma=0.21$, this dependence is not relevant for our purposes.

To obtain $\sigma(Q)$ we use the standard BBKS power spectrum (Bardeen et al. 1986). We also checked other power spectra (e.g. Einseinstein \& Hu 1999), finding no significant differences in the final results, i.e.

\begin{eqnarray}
\sigma(Q) \equiv \sigma(Q,\Gamma) \simeq \sigma_8 A(\Gamma) Q^{(-B(\Gamma)-C(\Gamma)Q)} \nonumber\\
A(\Gamma) \equiv 2.01 + 3.9 \Gamma ; B(\Gamma) \equiv 0.2206 + 0.361 \Gamma^{1.5} \nonumber \\
C(\Gamma) \equiv 0.182 + 0.0411 ln(\Gamma)     \label{eq:eqA8}
\end{eqnarray}

This fit is valid for $Q\ga 3 \mpch$ and $0.1\ga\Gamma\ga0.5$.

To obtain $P^{\star}_n(r^{\star})$, i.e. the probability that a sphere of radius $r^{\star}$ in redshit space contains $n$ objects when placed at random within the distribution, we have to implement the replacements given in equations (16) and (17) in all expressions entering in eq. (A1) but for computational reasons we choose to follow an equivalent procedure whereby $r$ is replaced by:

\begin{equation}
r^{\star} \left( \frac{[1-{\rm VEL}(\delta)]^4 - 1}{-4 {\rm VEL}(\delta)}\right)^{1/3}    \label{eq:eqA9}
\end{equation}
where the function ${\rm VEL}(\delta)$ is defined so that the peculiar velocity, $V$, of mass element at distance $r$ from the center of a spherical mass concentration (or defect) enclosing actual density contrast $\delta$ is given by: 
\begin{equation}
V = H~r~ {\rm VEL}(\delta)   \label{eq:eqA10}  
\end{equation}
where $H$ is the Hubble constant at the time being considered.
Betancort-Rijo et al. (2006) showed that:

\begin{equation}
{\rm VEL}(\delta) = -\frac{1}{3} \frac{d~ln D(a)}{d~ln a} \frac{{\rm DELK}(\delta)}{1+\delta} \left( \frac{d}{d\delta} {\rm DELK}(\delta) \right)^{-1}   \label{eq:eqA11}
\end{equation}
$D(a)$ is the growth factor as a function of the expansion factor, $a$, and ${\rm DELK}(\delta)$ is the inverse function of ${\rm DELT}(\delta_l)$ 
(see Sheth \& Thormen 2002):

\begin{eqnarray}
{\rm DELK}(\delta) = \frac{\delta_c}{1.68647} ( 1.68647 - \frac{1.35}{(1+\delta)^{2/3}} - \nonumber \\
- \frac{1.12431}{(1+\delta)^{1/2}} + \frac{0.78785}{(1+\delta)^{0.58661}} )    \label{eq:eqA12}
\end{eqnarray}
$\delta_c$ is the linear density contrat for spherical collapse model, which for the concordance cosmology at present is 1.676. 
The logarithmic derivative of $D(a)$ is, for a given cosmology, a function of redshift. We can approximate it as:
\begin{equation}
\frac{d {\rm ln}D(a)}{d {\rm ln} a} \simeq 0.47 \left(\frac{(1+z)^{3}}{\Omega_{m}[(1+z)^{3}+\Omega_{\lambda}/\Omega_{m}]} \right)^{0.6}
\end{equation}

It must be noted that, although $r$ has to be replaced by eq. A2 in all its appearances in eq. \ref{eq:eqA9}, for computational reasons we only implement 
that replacement on $P(\delta_l,r)$ and in the explicit appearance of $r$ in eq. \ref{eq:eqA2}.

\section{The effect of cosmic variance}

In all the equations used in this work, the mean number density of objects were assumed to be obtained from an arbitrarily large sample. The question is which value of $\bar{n}$ must be used to obtain the expected theoretical value of voids within a specified sample: must we use the universal mean number density, when this is available from a much larger sample, or the mean within the sample under consideration? Conversely, if only the local mean is available, will it bias the expected number of voids within the sample?

Equation (29) gives (for samples at fixed $z$) the expected number of voids, $\bar{N}(r)$ (note that $\bar{N}(r)\equiv \bar{N}_L(r)$), within a sample of given shape and volume,  chosen randomly within a much larger volume with mean number density $\bar{n}$. $\bar{N}(r)$ can be also expressed in the form:

\begin{equation}
\bar{N}(r) = \int_{-\infty}^{\infty} \bar{N}(r)\mid\bar{n}' \left(\frac{e^ {-\frac{(\bar{n}'-\bar{n})^2}{2\sigma^2_{\bar{n}}}}}{\sqrt{2\pi}\sigma_{\bar{n}}}\right) d\bar{n}'   \label{eq:eqB1}
\end{equation}
$\bar{N}(r)\mid\bar{n}'$ represents the expected number of voids within the sample, conditional to having a mean number density within it equal to $\bar{n}'$. The parenthesis represents the probability distribution of $\bar{n}'$, which for large enough samples is simply a Gaussian with mean $\bar{n}$ and variance $\sigma_{\bar{n}}^{2}$ given by:

\begin{equation}
\sigma_{\bar{n}}^2 = \bar{\delta}^2 + \frac{1}{\bar{n}V_s} \label{eq:eqB2}
\end{equation}
where $\bar{\delta}^2$ is the variance of the density fluctuations within the sample, which for the large sample volumes, $V_{s}$, usually considered may be approximated 
by its linear value.

Equation (36) can be solved using the following \emph{ansatz}:

\begin{eqnarray}
\bar{N}(r)\mid\bar{n}' = \bar{N}(r)_{res}(\bar{n}=\beta\bar{n}');\nonumber\\
(\sigma(r)^2 \rightarrow \sigma^2(r) - \bar{\delta}^2)   \label{eq:eqB3}
\end{eqnarray}
where the right hand side is equation (\ref{eq:eq31}) considered as a function of $\bar{n}$ evaluated at $\beta\bar{n}'$, $\beta$ is a parameter to be determined. equation (\ref{eq:eq31}) should be 
evaluated using the constrained $\sigma(r)$ indicated in the last parenthesis. The reason for this is that within samples with a fixed value of $\bar{n}'$ the field of density contrast linearly extrapolated to the present behave, with high precision, like a uniform Gaussian field with a constrained power spectrum (so that on any scale $\sigma(r)'^2=\sigma(r)^2 - \bar{\delta}^2$, see Rubi\~no et al. 2008). This constraint on the power spectrum implies that at a given scale $r_0$ the value of $\sigma(r_0)$ is slightly different from the unconstrained one, and that the local shape of $\sigma(r)$ in the neighborhood of $r_0$ (for any $r_0$) also changes. As we saw before,  $\bar{n}_v(r_0)$ depends both on $\sigma(r_0)$ and on the local shape of $\sigma(r)$, quantified by $G_\gamma$. Therefore, we can write:

\begin{eqnarray}
\bar{N}(r_0)_{res}(\bar{n}=\beta\bar{n}') \simeq \nonumber\\
\left(1 - \frac{\bar{\delta}^2}{\sigma^2(r_0)}\right)\bar{N}(r_0)(\bar{n}=\beta\bar{n}')   \label{eq:eqB4}
\end{eqnarray}
We may also use the approximation:

\begin{equation}
\bar{N}(r_0)(\bar{n}') \simeq \bar{N}(r_0)(\bar{n}_0)\left(\frac{\bar{n}'}{\bar{n}_0}\right)^{G_{\bar{n}}}  \label{eq:eqB5}
\end{equation}

The values for $G_{\gamma}$ ($\equiv$ $G(r,\gamma)$), $G_{\bar{n}}$ ($\equiv G(r,\bar{n})$, see equation (\ref{eq:eq38}) for a general definition) are given in section (\ref{res}).
Using expressions (B3) and (B4) in (B5) we find, after eliminating $\bar{N}(r)$:

\begin{equation}
1 = \left(1 + \frac{\bar{\delta}^2}{\sigma^2(r_0)}\right)^{G_{\sigma}-G_{\gamma}} \left(1+\frac{G_{\bar{n}}(G_{\bar{n}}-1)}{2}\sigma_n^2\right) \beta^{G_{\bar{n}}} \label{eq:eqB6}
\end{equation}
Therefore, for $\bar{\delta}^2 << \sigma^2(r_0)$:

\begin{equation}
\beta \simeq 1 -  \left( (G_{\sigma} - G_{\gamma})\frac{\bar{\delta}^2}{\sigma^2(r_0)} + \frac{G_{\bar{n}}(G_{\bar{n}}-1)}{2} \sigma_n^2 \right) \frac{1}{G_{\bar{n}}} \label{eq:eqB7}
\end{equation}
For most relevant voids $G_{\gamma}$, $G_{\bar{n}}$ are close to 1, and for all the samples that we shall consider $\bar{\delta}^{2}\la10^{-2}$. Therefore, $\beta$ is within 1\% or 2\% percent from unity, implying that:

\begin{equation}
\bar{N}(r)\mid\bar{n}' \simeq \bar{N}(r)(\bar{n}')  \label{eq:eqB8}
\end{equation}
to a very high degree of accuracy. That is, the conditional expected number of voids within a sample with certain density $\bar{n}'$ is the same as the expected number of voids within a random sample with the same shape and volume chosen at random within a much larger volume with mean density $\bar{n'}$. For very small samples, however, we would be underestimating the conditional number of voids if we use the unconditional expected number corresponding to the actual number density within the sample.

It is the conditional expected number of voids that we use to run tests for cosmological parameters because it leads to a somewhat more restrictive test (it uses more information). Furthermore, the constraint on $\sigma_8$ coming out from a test based on values of $\bar{n}'$ within the samples may be combined with that coming from tests using voids statistics as if they were independent tests. 
It must be noted, however, that a subdivision of the sample into smaller samples to use the conditional void statistics to conduct the test may lead to a further small increase of the efficiency of the test but only in as much as $\beta$ remains very close to 1 for the subsamples.

\section{Tables with the comparison between our formalism and the Millennium Run}

In this Appendix we provide the tables with the values corresponding to the figures of section \ref{res} which allows a more quantitative comparison of the results found using our formalism to with those found in the Millennium Run. We also present numbers for other underdense statistics as well. 

\begin{table*}
\begin{tabular}{|c|cccccccccc}
\hline
Radius $[\mpch]$  &  n=0 ($sim$) & 0 ($f$)&  1 ($sim$)  & 1 ($f$)&  2 ($sim$)  & 2 ($f$)&  3 ($sim$)  & 3 ($f$)&  4 ($sim$)  & 4 ($f$) \\
\hline
9  & 2.5600 $\pm$ 0.2300 & 2.5150 & 3.0170 & 3.1540 & 3.3180 & 3.5330 & 3.5730 & 3.7640 & 3.7530 & 3.9010 \\
10 & 1.1240 $\pm$ 0.1380 & 1.0870 & 1.5340 & 1.5630 & 1.8160 & 1.9050 & 2.0770 & 2.1730 & 2.2770 & 2.1730 \\
11 & 0.4568 $\pm$ 0.0970 & 0.4390 & 0.7228 & 0.7123 & 0.9196 & 0.9420 & 1.1040 & 1.1420 & 1.2540 & 1.3210 \\
12 & 0.1705 $\pm$ 0.0378 & 0.1654 & 0.3144 & 0.3008 & 0.4266 & 0.4281 & 0.5334 & 0.5504 & 0.6374 & 0.6681 \\
13 & 0.0561 $\pm$ 0.0095 & 0.0581 & 0.1246 & 0.1173 & 0.1884 & 0.1796 & 0.2407 & 0.2440 & 0.2971 & 0.3103 \\
14 & 0.0176 $\pm$ 0.0038 & 0.0190 & 0.0436 & 0.0462 & 0.0765 & 0.0695 & 0.1026 & 0.0998 & 0.1300 & 0.1327 \\
15 & 0.0043 $\pm$ 0.0013 & 0.0058 & 0.0137 & 0.0156 & 0.0269 & 0.0249 & 0.0409 & 0.0376 & 0.0528 & 0.0523 \\
\hline
\end{tabular}
\caption{$P_{n}(r)$ for dark matter halos with masses larger than $6.6 \times 10^{11} \Msunh$ in real space. The number density of
these halos is $5 \times 10^{-3} (\mpch)^{-3}$. Here we present results for the range n=0 to 4 for the Millennium Run (Sim) and the corresponding predictions obtained using our formalism ($f$). Note that all $P_{n}$ values are in units of $10^{-2}$.}
\end{table*}

\begin{table}
\begin{tabular}{|c|cc|}
\hline
Radius $[\mpch]$   & $P^{f}_0$  &  $P^{sim}_{0}$  \\
 \hline
2 &    $8.8600$   &    $8.8670 \pm 0.01241 $ \\
3 &    $7.6800$   &    $7.2860 \pm 0.02430 $ \\
4 &    $5.6000$   &    $5.4760 \pm 0.03230 $ \\
5 &    $3.9200$   &    $3.7860 \pm 0.03610 $ \\
6 &    $2.4400$   &    $2.4240 \pm 0.03540 $ \\
7 &    $1.4400$   &    $1.4400 \pm 0.03200 $ \\
8 &    $0.7920$   &    $0.8020 \pm 0.02700 $ \\
9 &    $0.4123$   &    $0.4177 \pm 0.02020 $ \\
10&    $0.2034$   &    $0.2041 \pm 0.01400 $ \\
11&    $0.0951$   &    $0.0936 \pm 0.00873 $ \\
12&    $0.0422$   &    $0.0403 \pm 0.00500 $ \\
13&    $0.0178$   &    $0.0161 \pm 0.00271 $ \\
14&    $0.0071$   &    $0.0060 \pm 0.00132 $ \\
15&    $0.0027$   &    $0.0019 \pm 0.00059 $ \\
\hline
\end{tabular}
\caption{Comparison for the VPF ($P_0$) in a (redshift space) mock galaxy distribution obtained using our formalism (2nd column, $f$) and numerical simulations (3rd column, $sim$). The error is the {\it rms} over an ensemble of 60 mock catalogs. These results were computed for galaxies brighter than $M_{r}= -20.4+5logh$. All the values of the VPF are given in units of $10^{-1}$.}
\end{table}

\begin{table*}
\begin{tabular}{|c|cccc|}
\hline
Radius $[\mpch]$  & $P^{f}_{1}$ &  $P^{sim}_{1}$   &  $P^{f}_{4}$   & $P^{sim}_{4}$      \\
\hline
8  &   $7.0660$ & $7.2540 \pm 0.124  $ &  $5.8540  $  &  $6.0260 \pm0.042  $\\
9  &   $4.2350$ & $4.4570 \pm 0.126  $ &  $4.2750  $  &  $4.6270 \pm0.051  $\\
10 &   $2.3700$ & $2.5130 \pm 0.155  $ &  $2.8520  $  &  $3.1870 \pm0.060  $\\
11 &   $1.2430$ & $1.3220 \pm 0.083  $ &  $1.7600  $  &  $1.9870 \pm0.059  $\\
12 &   $0.6137$ & $0.6477 \pm 0.054  $ &  $1.0120  $  &  $1.1440 \pm0.053  $\\
13 &   $0.2853$ & $0.2936 \pm 0.032  $ &  $0.5451  $  &  $0.6053 \pm0.004  $\\
14 &   $0.1251$ & $0.1244 \pm 0.031  $ &  $0.2755  $  &  $0.2963 \pm0.029  $\\
15 &   $0.0517$ & $0.0500 \pm 0.011  $ &  $0.1307  $  &  $0.1360 \pm0.017  $\\
\hline
\end{tabular}
\caption{$P_1$ and $P_4$ for the same galaxy distribution as in Table 2. The superscripts {\it f} and {\it sim} denote results obtained using our formalism and numerical simulations respectively. All $P_{n}$ are given in units of $10^{-2}$.}
\end{table*}

\begin{table*}
\begin{tabular}{|c|cccc|}
\hline
Radius $[\mpch]$  &  $\bar{n}^{f}_{v}$ (halos)   &  $\bar{n}^{sim}_{v}$ (halos)  &  $\bar{n}^{f}_{v}$ (gals)   & $\bar{n}^{sim}_{v}$ (gals)   \\
\hline
9  & 6.325 & 6.460 $\pm$ 0.090  & 6.318  & 6.591 $\pm$ 0.084 \\
10 & 3.377 & 3.348 $\pm$ 0.058  & 3.531	 & 3.471 $\pm$ 0.061 \\
11 & 1.795 & 1.906 $\pm$ 0.044  & 2.040  & 2.040 $\pm$ 0.047 \\
12 & 0.969 & 1.000 $\pm$ 0.032  & 1.093  & 1.120 $\pm$ 0.036 \\
13 & 0.511 & 0.510 $\pm$ 0.023  & 0.596  & 0.598 $\pm$ 0.026 \\
14 & 0.256 & 0.236 $\pm$ 0.016  & 0.301  & 0.289 $\pm$ 0.019 \\
15 & 0.113 & 0.103 $\pm$ 0.011  & 0.141  & 0.130 $\pm$ 0.010 \\
16 & 0.047 & 0.041 $\pm$ 0.007  & 0.059  & 0.051 $\pm$ 0.008 \\
\hline
\end{tabular}
\caption{Simulation results versus theoretical predictions for the number density of voids defined alternatively by halos and galaxies [both with number density $5\times 10^{-3}(\mpch)^{-3}$] in redshift space in the Millennium run at $z=0$. Number densities are given in units of $10^{-5} (\mpch)^{-3}$. See text for details.}
\end{table*}

\begin{table*}
\begin{tabular}{|c|cccc|}
\hline
Radius $[\mpch]$  &  $\bar{n}^{f}_{v}$ (halos)   &  $\bar{n}^{sim}_{v}$ (halos)  &  $\bar{n}^{f}_{v}$ (gals)   & $\bar{n}^{sim}_{v}$ (gals)   \\
\hline
9  & 6.200   & 6.700 $\pm$ 0.085  & 6.258   & 6.681 $\pm$ 0.085 \\
10 & 3.247   & 3.378 $\pm$ 0.060  & 3.298   & 3.429 $\pm$ 0.060 \\
11 & 1.693   & 1.907 $\pm$ 0.046  & 1.865   & 1.963 $\pm$ 0.045 \\
12 & 0.906   & 0.971 $\pm$ 0.033  & 0.973   & 0.998 $\pm$ 0.032 \\
13 & 0.464   & 0.465 $\pm$ 0.022  & 0.508   & 0.508 $\pm$ 0.024 \\
14 & 0.223   & 0.207 $\pm$ 0.015  & 0.238   & 0.232 $\pm$ 0.017 \\
15 & 0.092   & 0.082 $\pm$ 0.009  & 0.103   & 0.092 $\pm$ 0.011 \\
16 & 0.035   & 0.026 $\pm$ 0.006  & 0.038   & 0.032 $\pm$ 0.006 \\
\hline
\end{tabular}
\caption{The same as in Table 4 but for $z=1$. Again, number densities are given in units of $10^{-5} (\mpch)^{-3}$.}
\end{table*}


\end{document}